\documentclass{article}

\usepackage{arxiv}

\usepackage[utf8]{inputenc} 
\usepackage[T1]{fontenc}    
\usepackage{hyperref}       
\usepackage{url}            
\usepackage{booktabs}       
\usepackage{amsfonts}       
\usepackage{nicefrac}       
\usepackage{microtype}      
\usepackage{makecell}
\usepackage{siunitx}
\usepackage{caption}
\usepackage{subcaption}
\usepackage{multirow}
\usepackage{xr}
\usepackage{cleveref}
\usepackage{graphicx}

\newcommand\blfootnote[1]{%
  \begingroup
  \renewcommand\thefootnote{}\footnote{#1}%
  \addtocounter{footnote}{-1}%
  \endgroup
}
\pdfoutput=1

\title{Online Engagement with Retracted Articles: \\ Who, When, and How?}

\author{
  Henry K. Dambanemuya$^\ast$ \\
  Northwestern University\\
  Evanston, IL 60208 \\
  \texttt{hdambane@u.northwestern.edu} \\
   \And
 Rod Abhari$^\ast$ \\
  Northwestern University\\
  Evanston, IL 60208 \\
  \texttt{RodAbhari2025@u.northwestern.edu } \\
   \And
  Nicholas Vincent \\
  Northwestern University\\
  Evanston, IL 60208 \\
  \texttt{nickvincent@u.northwestern.edu} \\
   \And
 Em\H oke-\'Agnes Horv\'at \\
  Northwestern University\\
  Evanston, IL 60208 \\
  \texttt{a-horvat@northwestern.edu } \\
}

\begin{document}
\maketitle

\blfootnote{$^\ast$ Equal Contribution.}

\begin{abstract}
Retracted research discussed on social media can spread misinformation. Yet we lack an understanding of how retracted articles are mentioned by academic and non-academic users. This is especially relevant on Twitter due to the platform's prominent role in science communication. Here, we analyze the pre- and post-retraction differences in Twitter attention and engagement metrics for over 3,800 retracted English-language articles alongside comparable non-retracted articles. We subset these findings according to five user types detected by our supervised learning classifier: members of the public, academics, bots, science practitioners, and science communicators. We find that retracted articles receive greater user attention (tweet count) and engagement (likes, retweets, and replies) than non-retracted articles, especially among members of the public and bots, with the majority of user engagement happening before retraction. Our results highlight the prominent role of non-experts in discussions of retracted research and suggest an opportunity for social media platforms to contribute towards early detection of problematic scientific research online.
\end{abstract}

\keywords{retraction, science of science, social media, attention economy, misinformation}

\section{Introduction}

Online engagement with retracted research (i.e., scholarly articles whose scientific credibility has been revoked) can have unexpected consequences. Prior to retraction, engagement with retracted articles can help detect flawed research and facilitate its retraction \cite{haunschildCanTweetsBe2021}, whereas after the retraction, engagement with retracted articles can help circulate knowledge of the retraction. At the same time, engagement with retracted research may amplify the flawed claims of retracted articles and therefore spread misinformation\cite{marcus2018scientist, southwellDefiningMeasuringScientific2022}. For example, widespread media coverage of an article purporting a link between the Measles, Mumps, and Rubella (MMR) vaccine and autism \cite{wakefieldRETRACTEDIleallymphoidnodularHyperplasia1998} prompted a measurable increase in vaccine hesitancy, which the World Health Organization recently named as one of the greatest threats to global health \cite{mottaQuantifyingEffectWakefield2021,butlerDiagnosingDeterminantsVaccine2015}. Despite the eventual retraction of this article, it continues to receive uncritical citations in the academic literature that undermine the intention of the retraction \cite{suelzerAssessmentCitationsRetracted2019, TimelineWakefieldRetraction2010}. 

The threat of misinformation from flawed scientific articles has taken on increased relevance at a time of heightened scientific distrust. The COVID-19 pandemic accelerated the decline of trust in science among key demographics, including American conservatives, who largely rejected public health efforts to curtail the coronavirus \cite{MajorDeclinesPublic2023}. At the same time, scientific articles were circulated broadly on social media sites, by those both supportive of and hostile to science \cite{beersSelectiveDeceptiveCitation2023}. Some of these articles would later get retracted. According to the Center for Scientific Integrity's Retraction Watch database, over 370 COVID-19-related articles have been retracted as of January 2023 \cite{RetractedCoronavirusCOVID192020}. Many of these articles received considerable online attention. For example, the Lancet's retracted study on Hydroxychloroquine was mentioned over 29,000 times on Twitter (now known as X), making it one of the top 5 most highly mentioned articles online in 2020 \cite{AltmetricTop100}. Although the article itself was retracted within one month of publication, it continues to be discussed online as proof that scientific institutions are incompetent or corrupt \cite{abhariWhenRetractionsFail2024}.

Recognizing these challenges, emerging research has begun measuring social media engagement with retracted research. These studies find that retracted studies are generally discussed more than non-retracted studies, particularly before their retraction \cite{pengDynamicsCrossplatformAttention2022, serghiouMediaSocialMedia2021}. Although existing studies reveal important trends in the aggregate, they have not looked at how different social media users engage with retracted research. The users of social media platforms are diverse, representing a range of scientific perspectives and backgrounds. This diversity enables a collective intelligence distinct from either academic publishing or journalistic attention. For instance, members of the public may share research articles for different reasons than academic users, and may therefore have distinct patterns of engagement with retracted research.  Accordingly, a one-size-fits-all approach limits the potential of social media platform creators to understand patterns of engagement with problematic research.

There is a need, therefore, to conduct a user analysis of online engagement with retracted research. To address this need, we measure online attention to retracted articles for the different types of users who post about them. We focus our analysis on Twitter due to the platform's prominent role in facilitating direct communication between scientists and the larger public \cite{cardona-grauIntroducingTwitterImpact2016}. While recent changes to the platform and its ownership have affected the platform's policies, the platform nevertheless remains one of the most prominent sites of engagement with scientific research. On this platform, \textit{we investigate how different users engage with retracted articles}. Specifically, we investigate the following questions:

\begin{itemize}
    \item RQ1: Does attention to and engagement with retracted articles on Twitter differ from non-retracted articles? If so:
    \item RQ2: How does attention to and engagement with (non)-retracted articles on Twitter differ across different user types?
    \item  RQ3: How does attention to and engagement with (non)-retracted articles on Twitter vary across different user types, before and after retraction? 
    \item RQ4: How does attention and engagement on Twitter vary with different keyword-related content?
\end{itemize}

To investigate these questions, we rely on the Retraction Watch database of retracted articles, Altmetrics's database of social media posts related to academic articles, and Twitter's research Application Programming Interface (API). These data sources enable us to obtain complete metadata associated with retracted research articles and social media posts that mention the research articles. With these data, we compute relevant measures and conduct a series of computational analysis to quantify and characterize user attention to and engagement with retracted research. Additionally, using user profile descriptions, we identify five types of users who interact with retracted articles, namely academics, science communicators, science practitioners, bots, and members of the public.

We find that retracted articles receive more attention and engagement than non-retracted articles. However, there is significant heterogeneity in Twitter attention to and engagement with (non)-retracted articles across different user types both before and after retraction. Thus, our work contextualizes the out-sized attention that retracted research receives on social media by showing that this attention is not evenly distributed across users. These findings shed light on both a weakness of our current information environment (i.e., the potential for retracted information to continue to be shared online) and a contention that might be untenable (i.e., the idea that misinformation attracts more attention than accurate information because of its inaccuracy)~\cite{vosoughi2018spread}.
 
Our work contributes to nascent research that examines user attention and engagement with (non)-retracted articles online~\cite{shemaRetractionsAltmetricBibliometric2019,pengDynamicsCrossplatformAttention2022}. We underscore the potential of the social media to foster critical discussions about problematic scientific findings across broad audiences both before and after retraction. At the same time, we highlight the diversity of users in retracted article discussions and the broad reach of retracted articles on digital media, especially with members of the public. Thus, we signify the importance of investing resources into addressing scientific misinformation online. To this end, our work contributes to the development of scalable tools to help investigate and understand different ways that diverse groups of online users engage with scientific research on social media. Finally, our findings invite considerations about how platform design choices can foster positive scientific exchanges and contribute towards early detection of problematic scientific research online.

\section{Related Work: Online Attention to Scientific Retractions}

Social media have emerged as key platforms for the public to access and discuss scientific information. On social media, science experts share and discuss science with experts and non-experts alike. However, the outcomes of these engagements have been mixed. For example, public scientists such as Neil deGrasse Tyson have been lauded for their online advocacy of key scientific issues as well as chastised for their public conduct while doing so (e.g., \cite{corneliussenNeilDeGrasseTyson2014}). One problem is that social media was never designed to communicate scientific nuance  \cite{malikIdentifyingPlatformEffects2000,dijckUnderstandingSocialMedia2013}. Additionally, in order to compete in the attention economy of social media, scientific information is often oversimplified or sensationalized  \cite{caulfieldConfrontingStemCell2016a}. Furthermore, individuals in polarized ``echo chamber'' environments judge the credibility of scientific information based on how well it conforms to their prior partisan beliefs \cite{delvicarioSpreadingMisinformationOnline2016a,druckmanEvidenceMotivatedReasoning2019}. These factors jointly contribute to making online social media hotbeds of scientific misinformation \cite{westMisinformationScience2021, southwellDefiningMeasuringScientific2022}. 

Retractions are one way of the crucial ways in which scientific publishers can address significant errors that originate within published research. While journal editors may issue statements of correction or expressions of concern in cases where the main findings of a paper are reliable, despite error(s), retractions are reserved for cases where the errors are significant enough to undermine the scientific integrity of the research \cite{barbourGuidelinesRetractingArticles2019}. Notably, it is observed that more than half of retractions are issued for reasons of scientific misconduct, including fabrication, falsification, and plagiarism \cite{fangMisconductAccountsMajority2012}. 

In theory, retractions allow the scientific community to correct scientific misinformation by repudiating the research from which they derive \cite{shebleNINEMisinformationScience2021}. However, this determination is based on the values of the scientific community, not necessarily those of social media users \cite{mertonSociologyScienceTheoretical1973a}. In academic writing, citing retracted research as credible is widely discouraged \cite{barbourGuidelinesRetractingArticles2019}. Yet, no clear norm exists for online mentions which may serve several functions including indicating agreement or disagreement, asking clarifying questions, or promoting the article. In fact, recent research suggests that retracted papers receive more online attention than non-retracted papers, even when excluding tweets that question the article or call for its retraction \cite{serghiouMediaSocialMedia2021}. A recent study of cross-platform attention to retractions further finds that most retractions are issued after popular attention to an article's findings have been exhausted, limiting its corrective potential \cite{pengDynamicsCrossplatformAttention2022}. 

Moreover, although retraction notices were designed to correct the public record, they can also give visibility to flawed research that was previously out of view. A recent study shows that retractions based on research misconduct attract substantially more online attention than retractions due to error \cite{shemaRetractionsAltmetricBibliometric2019}. However, it is important to consider how these patterns are affected by different types of users. For example, it's not clear if this attention is primarily from members of the public, who may be attracted to the controversy generated by the controversy, or if all users, including academic users, tweet more often about retracted research. 

Thus, a crucial step towards evaluating social media's role in correcting scientific misinformation is understanding how different users attend to retractions. Although Twitter is not representative of the global population or social media writ large, the platform hosts a variety of interests from both within and beyond academia. The platform's diversity of users and knowledge affords a collective intelligence that may help to identify errors or inconsistencies that might otherwise go unnoticed \cite{eveCaseMoreDiversity2021,nielsen2017gender}. For example, a recent study demonstrates how early critical discussions of two COVID-19 articles on Twitter detected problems that were later cited as reasons for retraction \cite{haunschildCanTweetsBe2021}. When problems in research findings are discovered, social media enables users to notify relevant entities, such as journal publishers, by mentioning or ``tweeting at'' the user accounts of these entities \cite{daneshjouResearchTechniquesMade2021}. In this way, social media have the capacity to both respond to and facilitate retractions.

At the same time, the vast majority of tweets that mention scientific research receive little to no follow-up engagement in the way of further replies and retweets. \cite{neigerEvaluatingSocialMedia2013}. For instance, research on non-retracted science finds that over 80\% of science-mentioning tweets were purely informational and expressed no discernible stance towards the mentioned article \cite{naUserMotivationsTweeting2015}. Additionally, informational tweets are largely produced by bot accounts affiliated with science organizations that exist to disseminate new science quickly and, typically, without commentary \cite{didegahInvestigatingQualityInteractions2018}. However, the presence of bots varies substantially by discipline, producing over 60\% of tweets in the natural sciences but only 20\% in the humanities and social sciences. Ultimately, challenges of scientific engagement on social media raise important questions not just about the amount of attention towards retracted research, but also about who pays attention to science. If retracted article discussions are limited to bot activity with no further engagement, then the corrective potential of social media will also be limited. 

Finally, recent research finds that retracted articles receive greater attention than non-retracted articles before their retraction, and that post-retraction attention is significantly less for popular retracted articles \cite{serghiouMediaSocialMedia2021}. Yet it is difficult to extrapolate from these findings how much post-retraction attention is responsive to the retraction itself. From a misinformation correction perspective, it is more concerning if post-retraction attention is unaware of the retraction than if the retraction is acknowledged. 

\section{Data \& Methods}

\subsection{Datasets: Altmetric and Retraction Watch}

To investigate how retracted research is discussed on Twitter, we collect 1) a set of tweets that mention retracted articles and 2) another set of tweets that mention a matched sample of non-retracted articles. 

\paragraph{Retracted Articles.} We first rely on Retraction Watch's retracted article database to identify retracted research articles. The Retraction Watch database provides a comprehensive database of retracted articles. We used this database to obtain article-level information about retracted articles, including the journal, year, and Digital Object Identifier (DOI) of each article. We also obtain the date of the retraction notice and the reason for retraction \cite{RetractionWatchDatabasea}. Then we rely on Altmetric's database of social media posts to obtain unique identifiers of online references to academic articles. The Altmetric data contains tweets published between June 6, 2011 and October 8, 2019. From these data, we collect all Twitter IDs for tweets that mention a retracted article recorded in the Retract Watch database \cite{Altmetrics2015}. Using the Twitter IDs obtained from Altmetric, we query the Twitter API to obtain complete tweet-level metadata. The metadata includes the author id and the content of each publicly available tweet as of September 2021. We obtain complete tweet information for tweets mentioning 3,847 articles, or 56\% of the 6,868 retracted articles identified in the Retraction Watch data set. Of the 66,447 tweets identified, 657 included more than one article link in the tweet text, and thus were counted according to the number of unique mentions of articles in our data. With these tweets accounted for, our final sample contained 67,124 tweets about retracted articles.

\paragraph{Non-Retracted Articles.}To establish a baseline with which to compare retracted articles, we applied the same tweet retrieval method described above to a control set of non-retracted articles from a recent study on retractions \cite{pengDynamicsCrossplatformAttention2022}. Control articles were chosen with the following matching procedure: for each retracted article, the authors searched for a non-retracted article from the same journal with a comparable number of tweets.
Of these, 96.6\% of retracted and control article pairs had the same number of tweets $\pm1$. Of the 2,288 matched control articles, we were able to obtain complete tweet information for 2,085 articles with an aggregate total of 25,997 tweets. These data allow us to comprehensively study Twitter's role in the discussions around retracted research in comparison to an appropriately chosen baseline of non-retracted articles.

\paragraph{Attention \& Engagement.} We use Twitter's API to get detailed metrics about the number of tweets as well as the number of times a tweet was ``liked'', ``retweeted'', or received a ``reply''. These metrics form the basis of two types of measures: Twitter attention (i.e., number of tweets per article, retweets not included) and engagement (i.e., the number of likes, retweets, and replies that a tweet receives) . Thus, we measure attention by aggregating the count of all tweets except retweets. We exclude retweets as a measure of attention because they simply repeat the original tweet, generally for the purpose of boosting the original message. Thus, retweets are considered a measure of engagement rather than attention. Thus, we measure Twitter engagement aggregating the counts of likes, retweets, or replies to a given tweet.

\paragraph{Language.} 74\% of our tweets are written in English. The remaining tweets cover a wide variety of languages. For attention and engagement analysis, we considered all tweets; for content analysis, we included only tweets that contained English keywords.

\paragraph{Ethics Statement.} Our research team exercised an abundance of caution when working with these digital trace data. We designed our research methods in accordance with our institution’s Institutional Review Board (IRB) and the norms expressed within the Association of Internet Researchers Ethical Guidelines \cite{Franzke}. From the Twitter API, we collected user information for the purpose of identifying engagement and user types and only report results at the aggregate level. We additionally maintained users’ privacy preferences by only collecting data from public accounts. Any user information that could possibly be identifying, namely the users’ Twitter ID, their profile description, and the content of their shared tweets, has been stored within a secure cloud environment.

\subsection{User Type Classification}
\label{sec:user-type-classification}

To classify user types, we rely on a combination of manual, rule-based, and machine learning approaches on 26,260 English user descriptions. 20,195 descriptions are of users who mentioned retracted articles and 6,065 descriptions are of users who mentioned non-retracted articles. To manually label user types based on their descriptions, we developed an initial code-book with five user types (i.e., members of the public, bots, science communicators, academics, and practitioners). These categories are mutually exclusive and represent a user's primary role. The categories are inspired by Altmetric's aggregate statistics about users who disseminate scientific articles online. One annotator labeled a set of 1,757 descriptions from users that tweeted about retracted articles and 525 descriptions from users that tweeted about non-retracted articles. To test the reliability of these labels, a second annotator coded 19.4\% of the data. Comparing the labels on this subset resulted in a Cohen's score of $\kappa=0.789$, which indicates good agreement. 

\begin{figure}[!h]
    \centering
    \includegraphics[scale=.5]{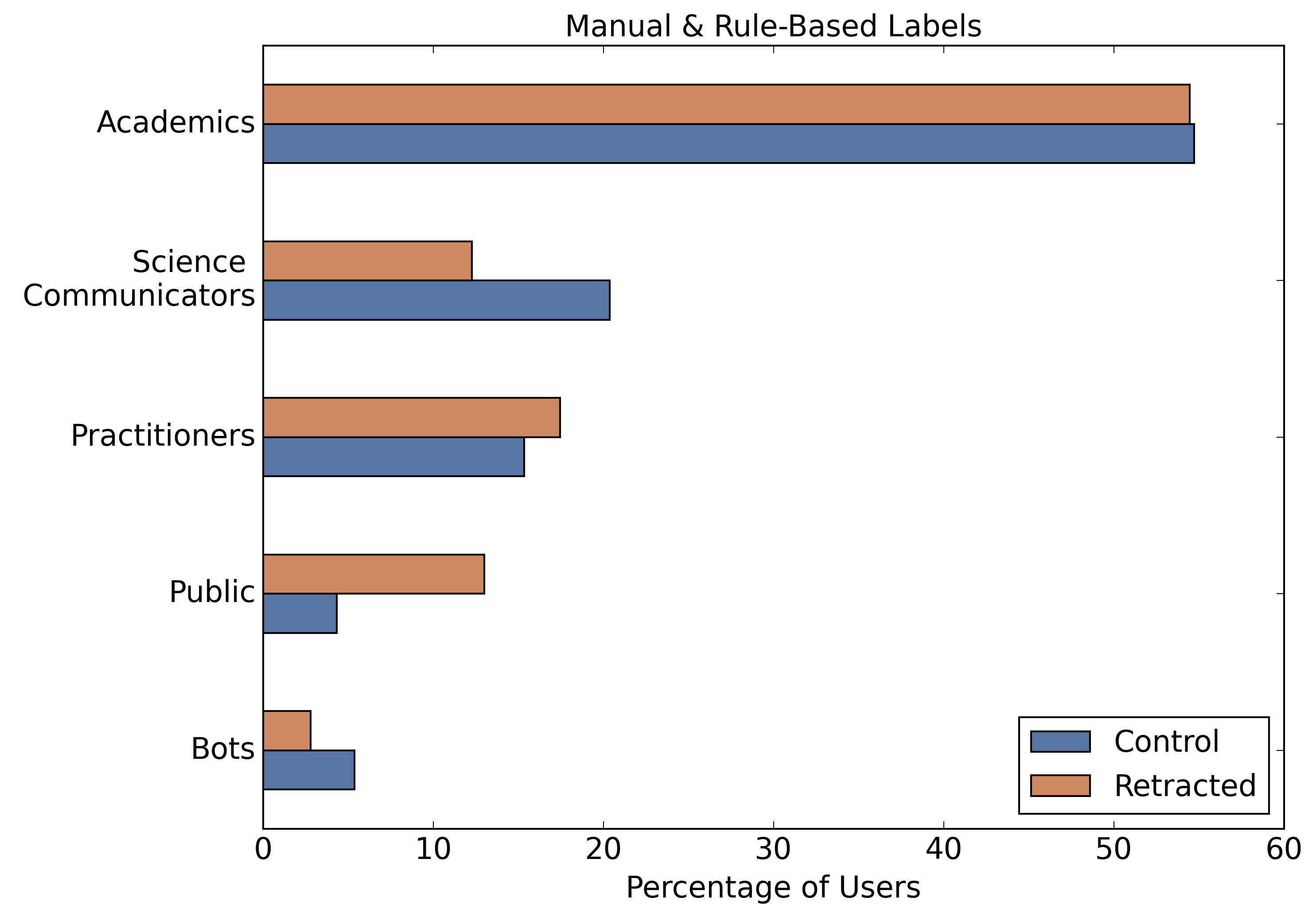}
    \caption{Percentage of Twitter users in the retracted (N=9,514) and non-retracted (N=2,495) groups that belong to each user type and identified through human annotation and a rule-based algorithm.}
    \label{fig:user_type_counts}
\end{figure}

Additionally, we used a set of keywords associated with each user type to construct a rule-based model that assigns users types to descriptions that contain any one of the keywords associated with the user type (Table~\ref{tab:user_type_keywords}). The keywords, determined a priori, are among the most frequently observed keywords in the set of manually-labeled user descriptions. Instead of including the full set of frequently used keywords for each user type, we rely on conservative keyword lists so as to maximize the true positive labels (i.e., user types that are correctly assigned by the rule-based method) and minimize false positive labels (i.e., user types that are incorrectly assigned). This process yields 7,757 rule-based labels for user types in the retracted group and 1,970 rule-based labels for user types in the non-retracted (i.e., control) group. Figure~\ref{fig:user_type_counts} provides a summary of the percentage of users that belong to each one of the five user types for the retracted and control groups. 

\begin{table}[!h]
\centering
\caption{List of keywords used to identify user types in rule-based classifier.}
\begin{tabular}{l|l}
\hline
\multicolumn{1}{c|}{User Type} & \multicolumn{1}{c}{Keywords} \\ \hline
Academics & principal investigator, prof, professor, researcher, scholar, scientist \\ \hline
Science Communicators & blogger, editor, journalist, writer \\ \hline
Health Practitioners & dentist, dietitian, doctor, md, nurse, nutritionist, physician, \\ & physical therapist, surgeon \\ \hline
Bots & bot, humanoid, automaton \\ \hline
Members of the Public & Based on manual annotations. Includes all users that  clearly do \\ & not belong to any of the other categories \\ \hline
\end{tabular}
\label{tab:user_type_keywords}
\end{table}

After combining the human labeled and rule-based labeled descriptions (i.e., 9,514 user descriptions in the retracted group and 2,495 user descriptions in the control group), we implement and evaluate three supervised classifiers. Random Forests~\cite{breiman2001random}, Decision Trees~\cite{breiman2017classification}, and Logistic Regression are trained on the task of classifying five user types from a feature vector constructed from users’ Twitter profile description. Since a disproportionately larger number of academic users is identified through the rule-based algorithm (see Figure~\ref{fig:user_type_counts}), we under-sample this group to maintain class balance. To create the feature vectors, we first build a vocabulary of all the unique words from all user descriptions combined. To reduce noise and the size of the vocabulary, we pre-process the user descriptions by (1) converting all the words to lowercase, (2) removing numbers, white spaces, punctuation, and stop words, and (3) lemmatizing the remaining words to reduce their inflectional forms to a common base or dictionary form. Then we represent the pre-processed text using Term Frequency — Inverse Document Frequency (TF-IDF) feature vectors. Specifically, we utilize the TFIDFVectorizer implementation of the scikit-learn~\cite{pedregosa2011scikit} module in Python. For each feature vector, the TF-IDF score of a term ($t$) in a user description ($d$) is measured by Equation~\ref{eq:tf-idf} as follows:

\begin{equation}
    tf\_idf(t,d) = tf(t,d) * idf(t),
    \label{eq:tf-idf}
\end{equation}

where $tf(t,d)$ represents the number of times that a term ($t$) appears in a user description ($d$) and $idf(t)$ is measures by Equation~\ref{eq:idf}:

\begin{equation}
    idf(t) = log\frac{1 + n_d}{1 + df(d,t)} + 1,
    \label{eq:idf}
\end{equation}

where $n_d$ is the total number of user descriptions and $df(d,t)$ is the number of user descriptions that contain the term $t$. Therefore, each user description is represented by a finite-length TF-IDF feature vector. We could not provide labels for users with no descriptions since their feature vectors have all zeros. Therefore, we treat all users without a description as a single, unknown user type. While there is reason to believe that these are likely new users, bots, or less active users, we avoid making any broad claims about this group. From the feature vectors, we then evaluate each model’s performance in predicting the ground truth labels using out of sample tests. To perform out of sample tests, we use 5-fold cross validation and report the model accuracy as the number of correct predictions divided by the total number of predictions. Finally, from the learned model, we project the remaining user descriptions on the best-performing model.

Using TF-IDF feature vectors described above to represent the descriptions of \textit{users that tweet about retracted articles}, we achieve a user type classification accuracy of 0.86 with the Decision Trees (CART) classifier. The Random Forest (RF accuracy=0.85) and Logistic Regression classifiers (LR accuracy=0.82) have comparable performance. Across all user types but academics, the Decision Trees classifier has lower classification errors compared to the other two models. Thus, to summarize how the different user types' tweets varied in terms of attention (i.e., the number of original tweets), engagement (i.e. likes, retweets, and replies), time (i.e., before/after retraction) and mentions of retraction-related keywords, we rely on the Decision Trees inferences due to the model's relatively low misclassification rates across most user types. When classifying \textit{users that tweeted about non-retracted articles}, we observe similar results with Decision Trees (CART accuracy=0.82), Random Forest (RF accuracy=0.80) and Logistic Regression (LR accuracy=0.75) classifiers.

To measure Twitter activity across user types, we compare the means of attention and engagement across three categories: retraction status, time-frame relative to retraction, and keyword status of the tweet. 

\subsection{Retraction Status and Time-Frame}
\label{sec:status-time}

Retraction status is determined by matching article DOIs to the Retraction Watch article database. Because many articles are tweeted about before retraction, we compare the timestamp of each tweet to the Retraction Watch's `Date of Retraction' to determine whether the tweet was posted before or after retraction. In order to provide a baseline comparison for retracted article tweets, we also determine the retraction time-frame of tweets related to non-retracted articles. Because each control article has a matched retracted article, we first create a retraction reference date for each control article by calculating the amount of time between the publication and retraction date of a matched retracted article. Then we determine if a tweet of a non-retracted article was produced before or after the retraction reference date. 

\subsection{Keyword Analysis}
\label{sec:keyword-analysis}

To further understand the content of tweets about retracted articles, we analyze each tweet using an interpretable keyword look up process.
Specifically, we aim to identify tweets that are substantively discussing an article's retraction using retraction-related keywords. To identify a list of relevant keywords, we use the ``Reasons for Retraction'' recorded in the Retraction Watch data, as well as the specific keyword `retract.' The complete list of words is shown in Table~\ref{tab:keyword_analysis}. We omit the following keywords that are ambiguous: `data', `image', `legal', `salami', and `notice'. The goal of this classification is to be broad enough to capture the scope of retraction-related keywords while being precise enough to avoid words used in different contexts.

We validate our keyword filtering approach by first dividing the dataset into tweets that contain a retraction-related keyword and those that do not. We then sample 145 retraction-aware and 140 retraction-unaware English tweets, representing a 10\% and 1\% sample, respectively. Two human annotators subsequently created ground-truth labels by manually labeling tweets according to whether the tweet discusses the retraction or not. The two annotators had an initial inter-coder agreement of 95.4\% (272/285) and achieve consensus in the second round of coding. In comparison to this ground truth based on human evaluation, 135 of the 145 tweets that the keyword-based matching found to be retraction aware were true positives (positive predictive value of 93.1\%) and 133 of the 140 tweets that did not contain retraction aware keywords were true negatives (negative predictive value = 94.8\%). These results indicate that such a keyword filtering approach can provide a reasonable proxy for detecting retraction aware tweets. 

\begin{table}[!h]
    \centering
     \caption{List of keywords drawn from the Reasons for Retraction as labeled in the Retraction Watch database.}
    \begin{tabular}{c|c|c|c}
         retract & duplication & unreliable & falsification\\ investigation & objections & plagiarism & unresponsive\\ authorship & fake & peer review & contamination\\ error & approval & misconduct & forge\\ withdraw & irb & hoax & breach\\ ethic & sabotage & criminal
    \end{tabular}
   
    \label{tab:keyword_analysis}
\end{table}

As a proxy for substantive conversation about retractions, we measure the fraction of all tweets which mention any of the retraction-related keywords (hereafter, ``keyword tweets''). Additionally, we divide keyword tweets into pre- and post retraction attention. Finally, we measure the attention to keyword tweets by comparing their engagement counts with the engagement totals across our entire sample of tweets mentioning retracted articles (e.g., what fraction of all the likes in our dataset are attributable to tweets that mention the word ``retract''). This simple approach allows us to evaluate the possibility of using social media to support meaningful discussion around research articles.

\section{Results}

\subsection{RQ1: User attention and engagement with (non)-retracted articles}
\label{result:RQ1}

\paragraph{Retracted articles receive more attention (i.e., original tweets, excluding retweets and replies) compared to non-retracted articles.} On average, retracted articles receive 7.09 original tweets, compared to 5.64 original tweets for non-retracted articles (Table~\ref{tab:attention-engagement-summary}). This finding contributes to the growing body of evidence that shows that retracted papers receive more online attention compared to non-retracted articles as well as prominent studies that show that misinformation spreads more broadly than the truth, in online platforms~\cite{pengDynamicsCrossplatformAttention2022,vosoughi2018spread}.

\begin{table}[!h]
\centering
\caption{Differences in attention (i.e., original tweets) and engagement metrics (i.e, likes, retweets, and replies) between (non)-retraction related tweets. Results show the mean(standard deviation) for attention and engagement metrics for control and retracted papers. P-Values are based on Mann-Whitney U test of independence.}
\label{tab:attention-engagement-summary}
\begin{tabular}{|ll|lll|}
\hline
Measure    &   & Control & Retracted  & P-Value \\ \hline
Attention  & Tweets   & 5.64(29.86) & 7.09(46.99) & p<0.001    \\ \hline
Engagement & Likes    & 11.3(82.61) & 17.03(137.92) & p<0.001   \\
           & Retweets & 8.06(48.19) & 11.77(86.57) & p<0.001   \\
           & Replies  & 1.13(8.37) & 2.42(20.91) & p=0.25\\ \hline
\end{tabular}%
\end{table}

\begin{figure*}[!h]
    \centering
    \includegraphics[scale=.6]{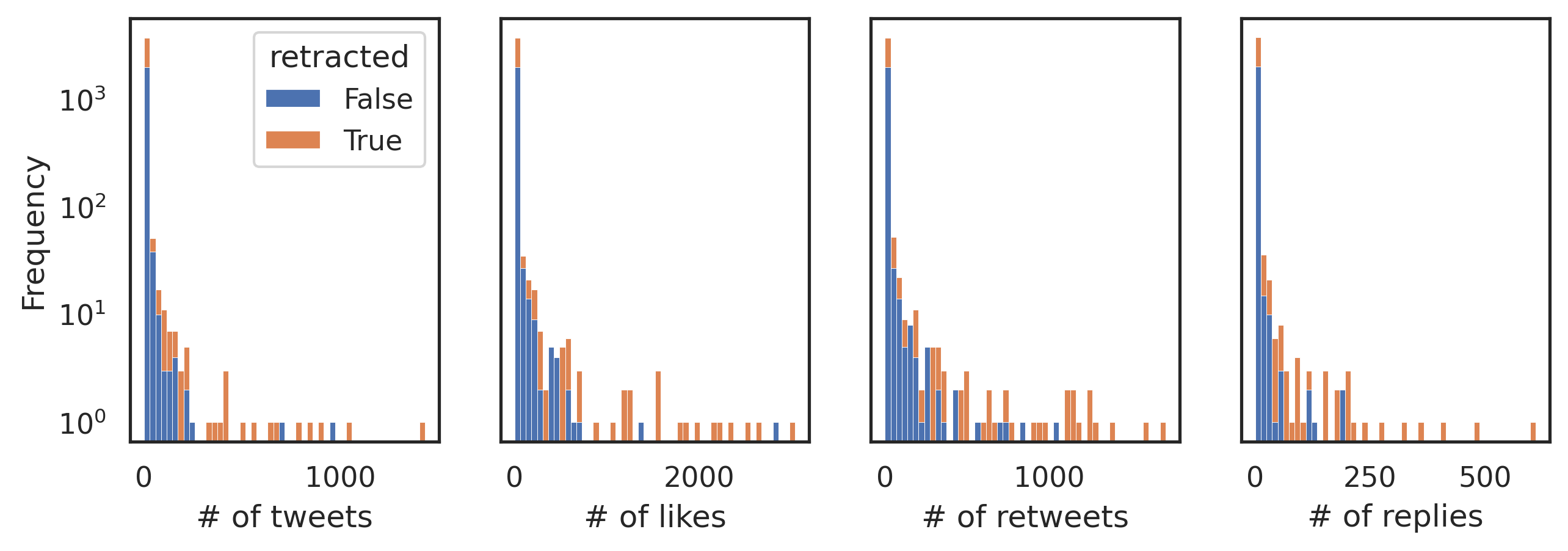}
    \caption{Distributions of Twitter attention and engagement metrics for retracted and control (non-retracted) articles.}
    \label{fig:distributions}
\end{figure*}

\paragraph{Retracted articles' tweets receive greater user engagement than those of non-retracted articles.} For each article, we measure user engagement by the average number of likes, retweets, and replies to all tweets associated with the article. Despite the engagement metrics having highly-skewed distributions (Figure~\ref{fig:distributions}), we observe that the engagement metrics are significantly (p<0.001) highly correlated with each other (Pearson correlation between like and retweet counts = 0.90, like and reply counts = 0.80, reply and retweet count = 0.81). Thus, in the following analysis, we rely on like counts as our engagement measure.

Mann-Whitney U tests of independence show significant differences in Twitter likes between retracted (17.03) and non-retraction (11.3) related tweets ($p<0.001$). We also observe significant differences in retweets between retracted (11.77) and non-retraction (8.06) related re-tweets ($p<0.001$). We observe no significant differences in replies for retracted and non-retraction related tweets (Table~\ref{tab:attention-engagement-summary}).

\subsection{RQ2: Attention and engagement with (non)-retracted articles across different user types}
\label{result:RQ2}

\begin{figure*}[!h]
    \centering
    \includegraphics[scale=.55]{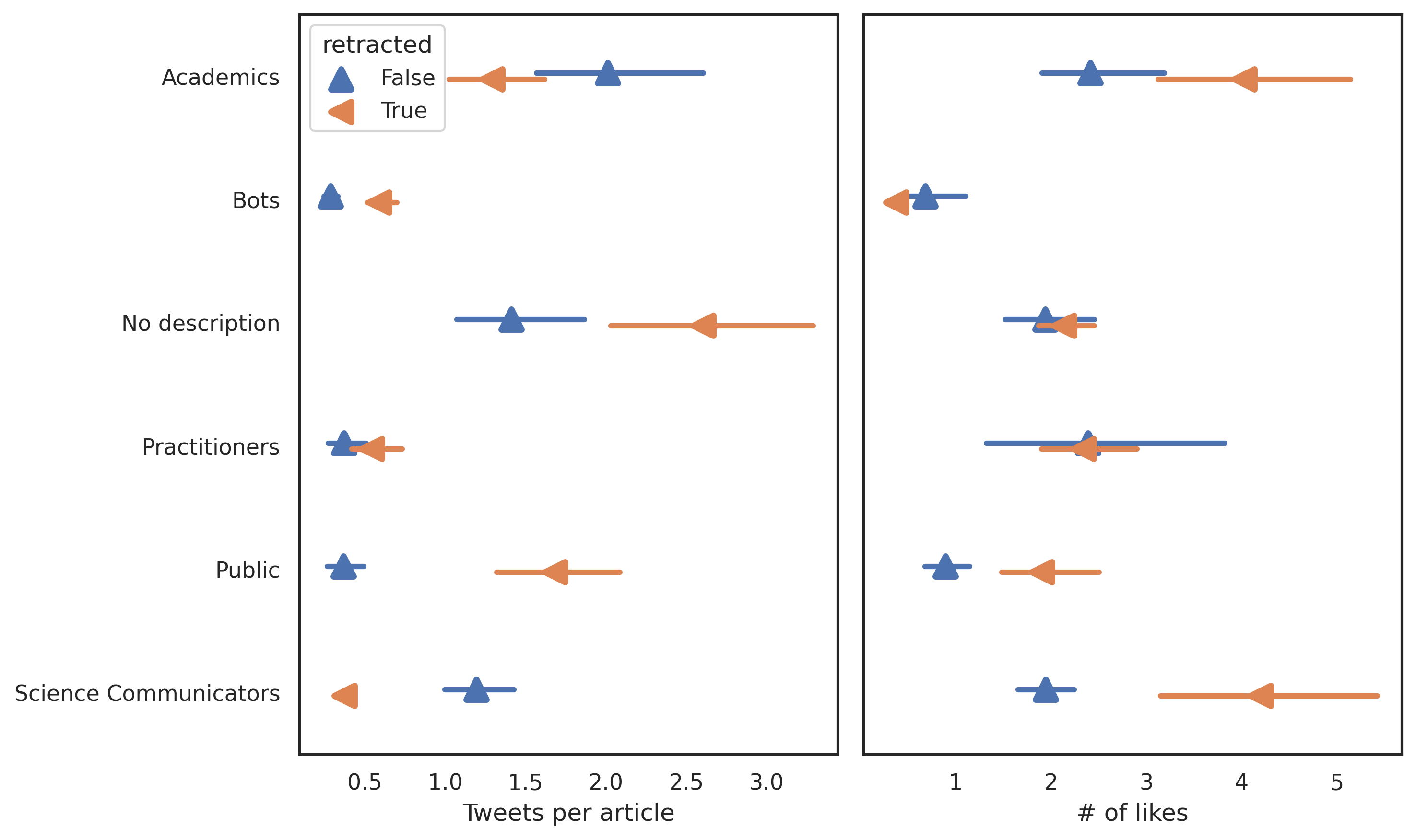}
    \caption{Mean and standard deviation (error bars) of user attention (tweets) and engagement (likes) with retracted and non-retracted articles by user types.}
    \label{fig:retracted_fig}
\end{figure*}

\paragraph{Out of the identifiable user types, public users are the most attentive to retracted articles.} Compared to other users, members of the public tweet the most about retracted articles, whereas academics tweet the most about non-retracted articles compared to other users (Figure~\ref{fig:retracted_fig}: Left). Among these two user groups, we observe significant heterogeneity in attention to (non)-retracted articles with public users more likely to tweet about retracted articles than academics and vice versa  ($\chi^{2}(1, N=16,211) = 2,331, p<0.001$). The finding that  members of the public tweet most about retracted research could be problematic because they are the least likely to be experts on the matter and are unlikely to encounter the retraction notice or other forms of corrections, compared to academics.

\paragraph{Academics and science communicators are most engaged with retraction-related tweets.} Although academics and science communicators tweet less about retracted articles compared to publics, they are the most engaged (in terms of like counts) with retraction-related tweets  (Figure~\ref{fig:retracted_fig}: Right). Taken together, these findings suggest that attention to and engagement with (non)-retracted articles is distributed unequally across different user types.

\subsection{RQ3: Attention and engagement before/after retraction}
\label{result:RQ3}

We further compare the proportion of Twitter attention and engagement with retracted articles before and after retraction (see Section~\ref{sec:status-time}). We use a baseline from a matching sample of non-retracted articles. 

\paragraph{Attention to retracted articles is high pre- instead of post-retraction}. Attention produced after retraction is considerably lower than before, with 30.57\% of post-retraction tweets receiving 31.43\% of total likes (Table~\ref{tab:time-tab}). This observation can be partly explained by the fact that articles are typically retracted after the attention cycle for an article's published findings has run its course \cite{pengDynamicsCrossplatformAttention2022}. 

Thus, to isolate the confounding effect of time on post-retraction attention, we compare post-retraction attention to a control set of non-retracted articles' tweets produced in the same period of time as matched retracted articles. For the non-retracted articles, we find that 27.00\% of tweets receive 38.01\% of likes produced in the time period after their matched sample's retraction. 

\paragraph{The relationship between attention and engagement with retracted articles is non-trivial.} We further observe that, although the number of post-retraction tweets is marginally higher (13.22\%) for retracted articles, the likes received on these tweets is lower for retracted articles (-17.31\%) (Table~\ref{tab:time-tab}). This finding suggests a non-straightforward relationship between attention to and engagement with retractions: retraction produces more original tweets than expected but fewer likes on these tweets. This nuances the finding that while, overall, retracted tweets receive more likes per tweets, the increase in likes is primarily concentrated on tweets produced before the retraction.  

\paragraph{Bot activity exhibits high attention to retractions, but low engagement with retraction-related tweets}. Bot accounts produced more than double the expected amount of post-retraction tweets for retracted articles (128.79\% increase), yet received half the expected likes on these tweets (-50\% decrease) (Table \ref{tab:time-tab}). This findings suggests that bots are uniquely responsive to the event of retraction, yet command unusually low levels of engagement on post-retraction tweets. 

\begin{table*}[ht]
\caption{Mean percentage of original tweets and likes contributed in the time period after retraction for each user. Percent change values show the degree of difference in original tweets and likes between retracted articles and non-retracted articles.}
\label{tab:time-tab}
\centering
\begin{tabular}{l|lll|lll|}
\cline{2-7} 
 & \multicolumn{3}{c|}{Post-Retraction Tweets}               & \multicolumn{3}{c|}{Likes on Post-Retraction Tweets} \\ \cline{2-7} 
                      & Control & Retracted & \% Difference & Control & Retracted & \% Difference \\ \hline
\multicolumn{1}{|l|}{Academics}             & 19.53\% & 19.57\%   & 0.20\%            & 29.98\% & 41.60\%   & 38.76\%           \\
\multicolumn{1}{|l|}{Bots}                  & 24.66\% & 56.42\%   & 128.79\%          & 45.71\% & 23.07\%   & -49.53\%          \\
\multicolumn{1}{|l|}{Practitioners}         & 22.72\% & 14.33\%   & -36.93\%          & 35.68\% & 13.67\%   & -61.69\%          \\
\multicolumn{1}{|l|}{Public}                & 19.25\% & 24.76\%   & 28.62\%           & 21.05\% & 30.43\%   & 44.56\%           \\
\multicolumn{1}{|l|}{Science Communicators} & 25.74\% & 17.52\%   & -31.93\%          & 31.53\% & 15.57\%   & -50.62\%          \\
\multicolumn{1}{|l|}{No Description}        & 23.14\% & 20.23\%   & -12.58\%          & 26.10\%  & 32.84\%   & 25.82\%         \\ \hline
\multicolumn{1}{|l|}{All Users} & \multicolumn{1}{l}{27.00\%} & 30.57\% & \textbf{13.22\%} & 38.01\%       & 31.43\%      & \textbf{-17.31\%}  \\   \hline
\end{tabular}
\end{table*}

\subsection{RQ4: Attention and engagement with keyword-related content}
\label{result:RQ4}

To further our understanding of the content of tweets about retracted articles, we use a keyword matching approach (see Section~\ref{sec:keyword-analysis}) to find tweets related to retracted articles, then we compare the proportions of keyword mentions to retraction status and time of retraction. 

\paragraph{Keyword-matching tweets are significantly more prominent for retracted than non-retracted articles.} As shown in in Table \ref{tab:keyword_tweets}, we observe approximately six times more tweets for retracted articles than non-retracted articles, and roughly seven times more likes per keyword tweet. Not entirely surprisingly, this finding confirms the relationship between keyword matches and retraction-related tweets. Furthermore, the proportion of keyword tweets increased significantly in the post-retraction window for retracted articles (717\%), whereas they increased only marginally in the post-retraction window for matched non-retracted articles ($\approx$50\%). More surprisingly, keyword tweets were roughly twice as frequent in the pre-retraction period for articles that would eventually be retracted than those which never were (2.42\% vs. 4.97\%). This suggests that retracted articles, overall, receive a higher than expected amount of retraction-related discussion before they are retracted. In the discussion section, we consider how platform design features could leverage this collective intelligence to predict questionable scientific content.  

\paragraph{Keyword-matching tweets are associated with high user engagement.} Although keyword-matching tweets make up a comparatively small proportion of tweets about retractions (13.36\%), these tweets receive out-sized engagement, accounting for 22.19\% of the total likes for tweets about retractions. Pre-retraction, keyword-mentioning tweets account only for 5\% of tweets but generate 8.74\% of likes. Post-retraction, keyword-mentioning tweets make up 40.60\% of tweets and account for 50.90\% of likes. The fact that the number of likes is higher than expected both prior to and following retraction suggests that Twitter users engage more with tweets that mention the possibility of retraction (pre-retraction), and the fact of the retraction (post-retraction). In sum, retracted articles do indeed see unique Twitter attention compared to control articles in the form of keyword tweets, and keyword tweets received additional engagement in form of likes.

\begin{table*}[ht]
\caption{Table showing the relative amounts of keyword tweets relative to non-keyword tweets by retraction status and time.}
\centering
\label{tab:keyword_tweets}
\begin{tabular}{l|lll|lll|}
\cline{2-7}
       & \multicolumn{3}{c|}{Keyword Tweets}  & \multicolumn{3}{c|}{Likes on Keyword Tweets} \\ \cline{2-7} 
       & Control & Retracted & \% Difference & Control    & Retracted    & \% Difference   \\ \hline
\multicolumn{1}{|l|}{Before} & 2.42\%  & 4.97\%    & 105.37\%       & 2.39\%     & 8.74\%       & 285.96\%         \\
\multicolumn{1}{|l|}{After}  & 3.64\%  & 40.60\%    & 1,015.38\%        & 4.92\%     & 50.90\%      & 934.55\%         \\ \hline
\multicolumn{1}{|l|}{Total}  & 2.68\%  & 13.36\%   & \textbf{398.51\%} & 3.14\%     & 22.19\%      & \textbf{606.69\%}  \\ \hline
\end{tabular}
\end{table*}

\begin{figure*}[ht]
    \centering
    \includegraphics[scale=.5]{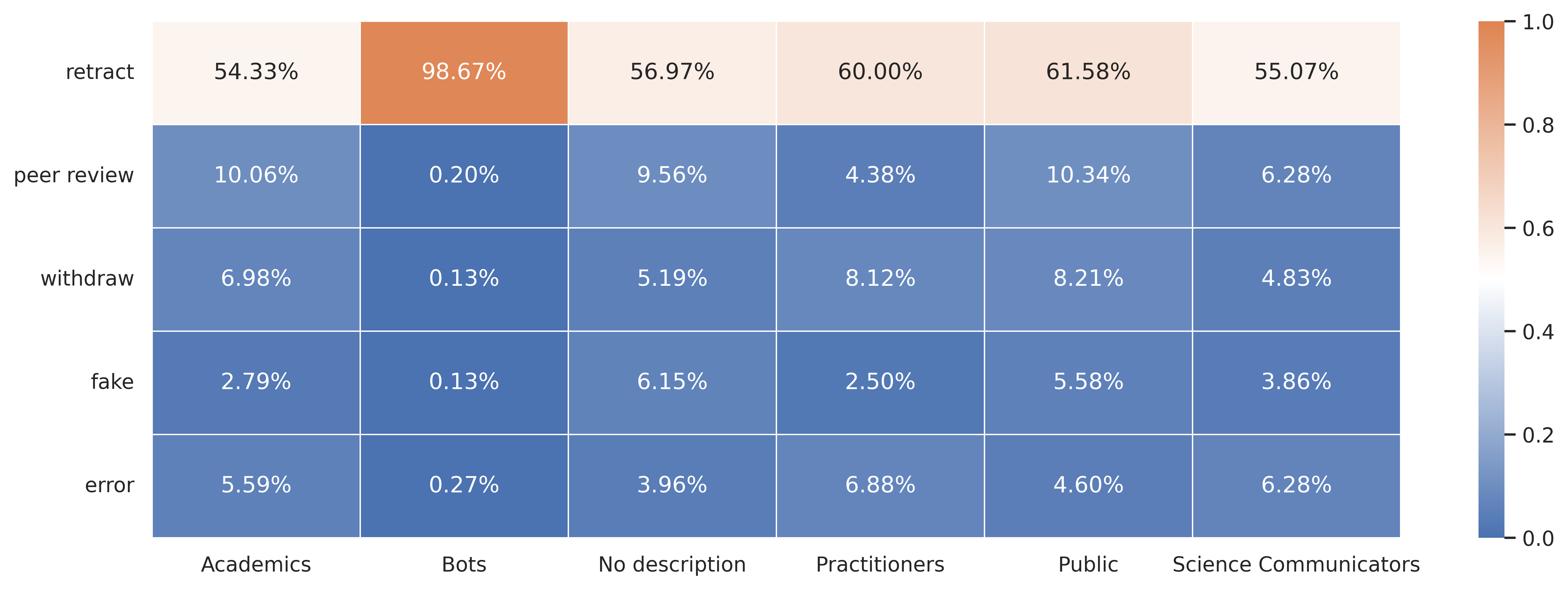}
    \label{fig:term_plot}
    \caption{Percentage of tweets that mention the top-5 retraction-related keywords broken down by user type. User types are inferred using a supervised decision tree classifier.}
\end{figure*}

\paragraph{Retraction-related keywords differ considerably among user types.} Figure~\ref{fig:term_plot} shows the user type breakdown of the number of tweets for the five most frequent keywords. The most mentioned-keyword, ``Retract,'' appeared in 73.2\% of keyword-containing tweets, while the next most common keyword, ``Peer Review,''  appears in only 5.8\% of keyword-containing tweets. Again, this is not unexpected given that reasons for retraction would be mentioned less frequently than the simple fact of retraction. What is surprising is the high prevalence of the keyword in bot related tweets. In fact, nearly all keyword mentions from bot tweets used the keyword ``Retract," with virtually no mentions of the other keywords. This, alongside the prior finding that bots were particularly attentive to articles after retraction, further implies that the singular purpose of many bots in calling attention to the fact of retraction. Other users had more even keyword distributions with less disparity. For human accounts, keywords related to reasons for retraction appear in 2-10\% of keyword tweets.

\section{Discussion}

Characterizing the nature of (science) discussion on social media is challenging. Such is the nature of online platforms that science-mentioning posts can appear to be more of an 'undigested dissemination' of information than a cogent discussion of issues~\cite{didegahInvestigatingQualityInteractions2018}. However, these discussions are intimately related to important issues in public health and science policy, and retractions are a key mechanism in the correction of scientific misinformation. Accordingly, it is essential to understand how distinct users engage with scientific articles before and after their retraction. 

One of the ways to make sense of the noise on social media is to consider trends in user activity. In doing so, we gain two primary insights. First, we observe pronounced differences between different users in how and when they tweet about retracted articles. We find that retracted articles receive proportionately more tweets from groups that are less represented in science, namely members of the public and bot accounts, as well as health practitioners. The opposite is true of academics and science communicators. These findings complicate the commonplace idea that academic discussion takes place primarily among academics on Twitter, e.g., on 'Academic Twitter.' The fact that public users produce the most attention to retracted articles prior to their retraction indicates that pre-retracted articles are of increased public interest even before their retraction. This contextualizes prior work by showing that the disproportionate attention that retractions receive on Twitter is at least partly due to the increased attention from public and bot users \cite{serghiouMediaSocialMedia2021, pengDynamicsCrossplatformAttention2022}. 

In addition, user types are represented differently depending on when they tweet relative to retraction. Notably, bots are the only group to tweet more after a retraction than before it. This supports previous research which finds that bots are prevalent in science communication on Twitter but departs from the characterization of their role as mere amplifiers of attention \cite{didegahInvestigatingQualityInteractions2018}. The fact that bots are more likely than other user types to mention retraction-related keywords in their tweets suggests that bots produce tweets that publicize an article's retraction. Although the poor overall engagement with bot tweets suggests that their relative impact is minimal, our research indicates that they play a unique role in the important task of publicizing retractions.

Furthermore, we find that academic users and science communicators receive high engagement in their tweets about retractions despite their relative sparsity in producing them. We learn that retracted articles receive different types of user activity compared to non-retracted articles, with significantly more original tweets, likes, and retweets compared to control articles. One possible reason for this discrepancy is that since retractions are inherently unusual, tweets about them generate increased engagement due to human curiosity. This is further supported by the fact that tweets specifically using retraction-related keywords received a larger proportionate share of likes and replies. Finally, we find that even in the case of retracted articles, credentialed experts receive more public engagement. 

\subsection{Social Media Engagement as a Bellwether for Science Communication}

Our findings represent an important contribution to the broader assessment of the role of social media as a forum for public science discussion. One of the implications of our research is that user engagement with science leaves behind signals that can be leveraged by scientific publishers and platform providers alike. The presence of retraction-related tweets before the occurrence of a retraction appears to indicate an undercurrent of critical discussion around some pre-retracted articles. Further support for this possibility is provided by the finding that non-retracted control articles are considerably less likely to mention retraction-related keywords, such as ``retract'', ``ethics'' and ``plagiarism,'' than pre-retraction articles. For articles with significant social media attention, establishing baseline values for the expected proportion of retraction-related tweets and comparing them with the observed proportions could contribute to approaches for detecting problematic sharing patterns. For instance, non-retracted articles receiving significantly higher than expected keyword-mentioning tweets may be a signal to academic publishers that an article may be objectionable. 

Conversely, retracted articles receiving significantly lower-than-expected portions of keyword-mentioning tweets after the retraction may indicate that the retraction itself has not been sufficiently communicated. In order to rectify this, social media platforms might take inspiration from scite\footnote{scite: \url{https://scite.ai/}}, a recently developed tool aimed at tracking citations of articles, including retracted ones \cite{nicholsonSciteSmartCitation2021}. This tool has already been incorporated into citation management software like Zotero\footnote{Zotero: \url{https://www.zotero.org/}}. Using scite, Zotero flags retracted articles and issues a visible warning when a user saves a retracted article. A similar feature could be added to social media platforms, either directly or through a browser plugin that lets users opt-in to receive such notifications. To the best of our knowledge, social media platforms such as Twitter do not plan to include any retraction notices, yet such platform features are essential given the extent to which academic research is shared and discussed on these platforms.

Crucially, social media platforms could monitor posts to ensure that retracted articles are mentioned correctly. One way this could be done is through integration with Twitter's Birdwatch program, now known as X's Community Notes. Such programs represent an attempt at organic content moderation by using crowdsourced labels to evaluate the accuracy of claims made on the platform. We believe that a similar approach could be applied to scientific findings. Domain experts vetted by Twitter could help review mentions of research advances in their respective disciplines to ensure that claims about retracted articles are accurate. 

\subsection{Limitations}
While user type and keyword analysis provide important context for the observed effects of retraction on Twitter engagement, these measures do not capture the full context of online conversations. The keywords used were selected from the ``Reasons for Retraction'' recorded in Retraction Watch's database. From this list, we took care to not include keywords that were too ambiguous (e.g., ``data'') and further broke our analysis down by keywords to confirm that this proxy analysis was informative. However, it is possible that some instances of the keywords were used in a niche context which we did not anticipate. Future work might develop a more elaborate list of keywords, or incorporate more expensive methods such as supervised learning.

In addition, since the TF-IDF model we used for classifying user types relies on the description text, we were unable to classify users who did not provide a profile description. Since understanding how different user types engage with retracted articles on Twitter is a critical and understudied component of engagement, future efforts could complement our findings by investigating potential ways to infer user types from user profile elements other than the description text e.g., users' membership lists, posting behaviour or social network structure. These approaches have previously been successful in identifying \emph{individual user types} on Twitter, e.g., bots~\cite{sayyadiharikandeh2020detection}, scientists~\cite{ke2017systematic}, or journalists~\cite{zeng2019detecting}, but were not fruitful in classifying \emph{multiple user types}, which was a central aim in our work.

When constructing a control group of non-retracted articles, we draw samples of articles from the same journals in which the retracted group of articles appeared. Then we identify non-retracted articles that matched the retracted articles in terms of the number of tweets associated with the article in question. While this seems like a reasonable proxy for the amount of promotion that the articles received, it is possible that articles from the same journal might receive the same number of promotional tweets if the journal promotes all articles equally. Hence, better methods could take into account the topic of the articles to account for the topical breakdown of retracted articles relative to non-retracted articles.

\section{Conclusion}
In today's political environment, science has become a topic of widespread contention. From COVID-19 to climate change, the solution to many of today's global problems depends in part on the dissemination of reliable scientific information. While social media has the potential to bring diverse constituents to scientific discussions in an unprecedented manner, it has also facilitated the spread of unreliable scientific information, including retracted research. In this context, social media platforms are called upon to create procedures that complement, rather than undermine, the retraction process. To support these efforts, our study has helped demystify Twitter engagement with retracted articles, indicating an association between tweet engagement, academic user status, and the retraction-related content of tweets. This opens up a window for further inquiry into how social media can be used for the development of research ideas, broadening participation in scientific discussions, and engaging with publishers and policy-makers.

Against a backdrop of science politicization, our research offers a note of cautious optimism, demonstrating how social media providers are capable of hosting relevant conversations about academic articles both before and after a retraction. Coupled with the timeliness and popular affordances of the platform relative to traditional forms of science communication, this supports the notion that social media can serve an important deliberative role within science. For instance, academics might use social media to debate whether they think an article is likely to be plagiarized or whether some data could be fraudulent. And science communicators may retweet these discussions to a broader audience than would otherwise be privy to the scientific deliberations. This suggests that social media platforms can help make science more transparent and more responsive to post-publication concerns. Ultimately, these results encourage design choices on social media that foster desirable scientific processes and incorporate more diverse views in scientific discussions.

\section*{Acknowledgments}
This work was supported by the U.S. National Science Foundation under Grant No. IIS-1943506. The authors would like to thank Hao Peng and Daniel Romero for providing the data and Herminio Bodon for labelling user type categories.

\bibliographystyle{unsrt}  
\bibliography{main}  

\end{document}